\documentclass[10pt, twocolumn]{IEEEtran}

\usepackage[dvips]{color}
\usepackage{epsf}
\usepackage{times}
\usepackage{epsfig}
\usepackage{graphicx}
\usepackage{epstopdf}
\usepackage{algorithm}
\usepackage{algorithmic}
\usepackage{amsmath}
\usepackage{amssymb}
\usepackage{amsxtra}
\usepackage{multirow}
\usepackage{mathtools}
\usepackage{cite}	
\usepackage{color}
\usepackage{bbm}

\DeclareMathOperator{\sign}{sign}

\title{Trajectory Optimization for Cooperative Dual-band UAV Swarms}

\author{
\IEEEauthorblockN{\large Hakim Ghazzai$^{1}$, Mahdi Ben Ghorbel$^{2}$, Andreas Kassler$^{3}$, and Md. Jahangir Hossain$^{2}$}\\
\IEEEauthorblockA{\small $^{1}$Stevens Institute of Technology, Hoboken, NJ, USA\\ Email: hghazzai@stevens.edu\\
$^{2}$The University of British Columbia, Okanagan Campus, Kelowna, BC, Canada\\ Email: \{mahdi.benghorbel, jahangir.hossain\}@ubc.ca\\
$^{3}$Karlstad University, Karlstad, Sweden\\ Email: andreas.kassler@kau.se\\
\vspace{-0.4cm}}
{\thanks {\vspace{-0.4cm}\hrule
\vspace{0.1cm} \indent Part of this work has been supported by the Knowledge Foundation Sweden through the SOCRA project.
This work was conducted when Hakim Ghazzai was with Karlstad University.\newline
2018 IEEE. Personal use of this material is permitted. Permission from
IEEE must be obtained for all other uses, in any current or future media,
including reprinting/republishing this material for advertising or promotional
purposes, creating new collective works, for resale or redistribution to servers
or lists, or reuse of any copyrighted component of this work in other works.
}
}
\vspace{-0.4cm}}

\begin{document}
\maketitle
\begin{abstract}
\boldmath
Unmanned aerial vehicles (UAVs) have gained a lot of popularity in diverse wireless communication fields.  They can act as high-altitude flying relays to support communications between ground nodes due to their ability to provide line-of-sight links. With the flourishing Internet of Things, several types of new applications are emerging. In this paper, we focus on bandwidth hungry and delay-tolerant applications where multiple pairs of transceivers require the support of UAVs to complete their transmissions. To do so, the UAVs have the possibility to employ two different bands namely the typical microwave and the high-rate millimeter wave bands. In this paper, we develop a generic framework to assign UAVs to supported transceivers and optimize their trajectories such that a weighted function of the total service time is minimized. Taking into account both the communication time needed to relay the message and the flying time of the UAVs, a mixed non-linear programming problem aiming at finding the stops at which the UAVs hover to forward the data to the receivers is formulated. An iterative approach is then developed to solve the problem. First, a mixed linear programming problem is optimally solved to determine the path of each available UAV. Then, a hierarchical iterative search is executed to enhance the UAV stops' locations and reduce the service time. The behavior of the UAVs and the benefits of the proposed framework are showcased for selected scenarios.
\end{abstract}\vspace{-0.2cm}
\begin{IEEEkeywords}
3D positioning and path planning, cooperative networks, mmWave communications, unmanned aerial vehicles.\vspace{-0.2cm}
\end{IEEEkeywords}
\section{Introduction}


Millimeter wave communications (mmWave), which exploits the frequency bands beyond $30$~GHz, represents a promising candidate for next generation mobile networks due to the under-utilization of this very large spectrum range and their ability to provide very high data rates~\cite{Du2017}. These properties are among the major challenges in 5G communications. First, the large band will allow to accommodate the exponentially increasing wireless devices with the emergence of Internet of things (IoT). Second, the need for higher data rates is surged by the increasing popularity of data-hungry wireless services, such as video streaming, cloud computing, online gaming, etc. However, one of the main limitations for mmWave is their requirement for short range line-of-sight (LoS) links~\cite{Rappaport2013}.

The development of small and lightweight unmanned aerial vehicles (UAV), aka drones, has gained increasing popularity in recent years. This has led to improved performances in terms of flight range, battery time, and weight in parallel to reducing production costs. These developments motivated their adoption for various commercial and civil applications to profit from the diverse advantages they present in terms of mobility, deployment flexibility, and remote or autonomous control. Examples of applications include traffic monitoring, border surveillance, disaster management, and  delivery services~\cite{7463007}. Inspired by their popularity in different domains, UAVs have attracted interest to adopt them for diverse wireless communication application. Specifically, profiting from their mobility, short-range line-of-sight (LoS) communication links can be established~\cite{PL3}. This can represent an ideal solution to enable mmWave communications. We propose then to use the UAV as a relay to enable a mmWave link between communicating devices. However, since UAVs are battery-powered, i.e., their energy is limited, efficient management of their operation is needed especially that a significant part of their energy is consumed for flight and hovering operations. Thus, it is of paramount importance to smartly plan their path such that they complete the required communication jobs while minimizing energy consumption and/or communication time.  
 
To the best knowledge of the authors, only few studies have focused on using drones with the mmWave band. Specifically, in~\cite{Rangan2014}, mmWave UAVs are considered as candidate for cellular networks. Some of the main challenges are discussed such as beamforming, blockage, and Doppler effect due to mobility. Hierarchical beam search and codebook design procedure is proposed to improve beamforming training performance and adaptive cruising is suggested to mitigate blockage but no detailed algorithms are provided. In~\cite{Kong2017}, UAVs are proposed to act as relays using mmWave communications to leverage effect of blockage. To enable accurate positioning that takes into account real channel gains, the authors proposed that the UAV samples the link qualities of mmWave beams while moving. Then, based on real-time sampling, channel estimation is improved using compressive sensing theory to gradually adjust its path. However, in these studies, path planning and energy constraints of the UAVs are not specifically considered.

Using and optimizing the operations of UAVs for various communication tasks have seen tremendous interest recently with the popularity of UAVs~\cite{Li2016,Fawaz2018,Jeong2018,Zhang2018,Fan2018}. 
But, the specific characteristics of mmWave communications in terms of channel model, LoS requirement, and throughput, make the presented solutions unsuitable for mmWave.
On the other hand, there are multiple studies on relaying optimization for mmWave communications using fixed or mobile relays but without employing UAVs~\cite{Wu2018,Deng2017,He2017}. 
However, the mobility and flexibility for UAVs provides additional degrees of freedom that should be investigated and optimized.

In this paper, we develop a generic optimization framework where multiple drones equipped with dual-band mmWave and microwave ($\mu$Wave) communication modules are employed to act as relays for multiple spatially distributed transceivers. Initially, located in a docking station, each drone is assigned to a selected set of transceivers such that their total service time is minimized. The service time accounts for the communication time needed to transmit the message and the flying time needed for the drones to arrive at destinations. To this end, we formulate a mixed integer non-linear programming problem (MINLP) that aims to determine i) the band to be used by each drone to support each pair of transceiver, ii) the locations at which the drones need to stop to relay the data, and iii) the path of each drone. This is performed while taking into account the speed of each drone, the energy capacity, the transmit power levels, and the locations of the transceivers that require support. UAVs employ the decode-and-forward (DF) strategy for data relaying. Its output can be easily extended to other relaying strategies. 

Due to the difficulty to determine an optimal solution of the MINLP, we develop a four-step iterative approach determining at each iteration optimized paths for the drones. In the first step, fixed UAV stops are optimized. Then, the MINLP problem is converted to a MILP and solved optimally. In the third step, a three-dimensional (3D) hierarchical search is developed to adjust the locations of the UAVs. Finally, the fourth step iterates between the second and the third steps. Selected simulation results are provided to highlight the performance of the proposed approach and illustrate the behavior of the UAVs versus the system parameters.

\section{System Model}
\label{Sec2}

\begin{figure}[t!]
\vspace{0.2cm}
  \centerline{\includegraphics[width=0.7\columnwidth]{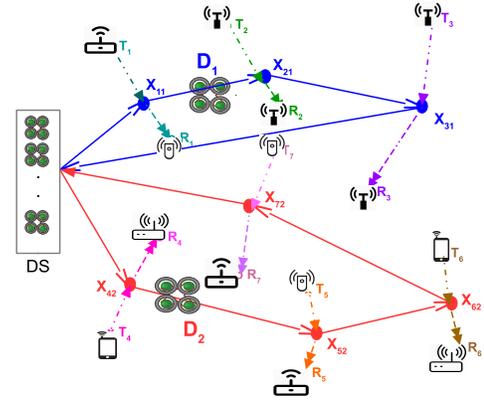}}\vspace{-0.3cm}
   \caption{\, Snapshot illustrating the system model. In this example, two UAVs are sent to support seven pairs of transceivers.    \normalsize}\label{Flowchart}
	\label{FigModel}\vspace{-0.4cm}
\end{figure}

We consider a UAV-based communication network located in a sub-region $\Omega \subseteq \mathbb R^3$, consisting of $D$ drones aiming to support the data transfer of $N$ pairs of transceivers. For each pair, we denote the transmitter by $T_n$ and the receiver by $R_n$. These nodes can be ground or aerial transceivers located at different positions identified by the 3D geographical coordinates $\boldsymbol{X}_{T{_n}}=(x_{T{_n}},y_{T{_n}},z_{T{_n}})$ and $\boldsymbol{X}_{R{_n}}=(x_{R{_n}},y_{R{_n}},z_{R{_n}})$, where $n=1,\dots, N$. Each pair of transceivers aims to transmit a message of size $M_n$ either via the traditional sub-3 GHz ($\mu$Wave) band or using the mmWave band (30 to 300 GHz). 
 Thereby, we consider that all nodes including the drones are equipped with two antennas: an omnidirectional antenna used for $\mu$Wave communications and a directive antenna with a gain $G$ used for mmWave communications. The drones will be employed as flying relays and must be exploited in an optimized manner to enhance the data transfer performance. 
The UAV cooperative scheme for multiple pairs of transceivers is illustrated in Fig.~\ref{FigModel}. 
We consider that $T_{n}$ transmits its signal with a constant transmit power equal to $P_{T_n}^{\text{$\mu$Wave}}$ and $P_{T_n}^{\text{mmWave}}$ depending on the used link. Similarly, each drone $d$, where $d=1,\dots,D$ will employ a fixed transmit power depending on the selected link that we denote by $P_{d}^{\text{$\mu$Wave}}$ and $P_{d}^{\text{mmWave}}$. Initially, the drones are assumed to be located at a docking station (DS) having as coordinates $\boldsymbol{X}_{DS}=\boldsymbol{X}_{0}=(x_{0},y_{0},z_{0})$. This position corresponds to the charging station to which the drones have to return back. 

\subsection{Channel Models}
As mentioned earlier, two radio bands are supported by the nodes to complete the data transmission: the $\mu$Wave band and the mmWave band. In this section, we present the channel models of each of the wireless links. We denote by $h_{U\rightarrow V}$ the channel gain representing the link between a node $U$ and another node $V$ where $U, V \in\{T_n,R_n, d\}$. In this study, the objective is to efficiently position the drones such that the transmission of all pairs of transceivers are enhanced. The overall transmission and flying times are relatively long compared to the channel coherence time and hence, we focus on the system performance based on their average statistics. 
Therefore, we only consider the large-scale path loss effect in the channel gain expressions that are expressed as follows:
\begin{equation}
h_{UV}(\Delta_{UV})=\frac{1}{\sqrt{PL_{UV}(\Delta_{UV})}},
\end{equation}
where $\Delta_{UV}$ is the distance separating the nodes $U$ and $V$ expressed as $\Delta_{UV}=||\boldsymbol{X}_U-\boldsymbol{X}_V||_2=\left((x_U-x_V)^2+(y_U-y_V)^2+(z_U-z_V)^2\right)^{\frac{1}{2}}$ where $||.||_2$ is the 2-norm distance. The term $PL_{UV}$ denotes the path loss effect. Finally, we denote by $\text{SINR}$ the signal-to-interference-plus-noise ratio, which is expressed as follows:
\begin{equation}
{\text{SINR}}_{UV}=\frac{P_UG_UG_V|h_{UV}|^2}{\mathcal I+N_0 B},
\end{equation}
where $P_U$ is the transmit power of the transmitter $U$, $G_U$ and $G_V$ are the antenna gains of the transmitter and receiver, $\mathcal I$ accounts the average external interference, $N_0$ is the power per frequency unit of an additive white Gaussian noise, and $B$ is the channel bandwidth. The interference effect over the mmWave band is ignored as all the nodes are using directional antennas~\cite{Rangan2014}. However, for the $\mu$Wave band, we consider that the drones using this band employ orthogonal transmission scheme such that the total $\mu$Wave bandwidth is equally divided among the drones that are using it. Hence, we denote by $\mathcal N^\text{$\mu$Wave}$ the number of drones using the $\mu$Wave band. In the sequel, we add the superscript $\lq\lq."^{\text{$\mu$Wave}}$ or $\lq\lq."^{\text{mmWave}}$ to distinguish between the different links.
\subsubsection{Path Loss Model for the $\mu$Wave Band}
As the studied system may include ground and flying transceivers, different channel gains are considered according to the types of the link. For instance, LoS links are considered between two flying nodes, e.g., one of the pair of transceivers and the drone, while non line-of-sight (NLoS) links are assumed between two ground nodes. However, for the link between a flying node and a ground node, LoS is supposed to be available with a certain probability related essentially to the altitude of the flying node~\cite{PL3}. Hence, the free space path loss effect in dB for a ground to ground (G2G) link is given as follows:
\begin{equation}
PL_{UV,\text{G2G}}^{\text{$\mu$Wave}}\hspace{-0.5mm}=\hspace{-0.5mm}PL_{UV}^{\text{NLoS}}\hspace{-0.5mm}=\hspace{-0.5mm}10n\log_{10}\left(\frac{4\pi f \Delta_{UV}}{C}\right)+L_{\text{NLoS}},
\end{equation}
where $n$ is the path loss exponent, $f$ is the carrier frequency, $C$ is the speed of light, and $L_{\text{NLoS}}$ is the average additional loss due to non-LoS link. Its value depends on the environment.\\
For an air-to-air (A2A) link, the path loss in dB is given as:
\begin{equation}
PL_{UV,\text{A2A}}^{\text{$\mu$Wave}}=PL_{UV}^{\text{LoS}}=10n\log_{10}\left(\frac{4\pi f \Delta_{UV}}{C}\right)+L_{\text{LoS}},
\end{equation}
where $L_{\text{LoS}}$ is the average additional loss due to LoS link.
Finally, for an air-to-ground (A2G), the average path loss effect is expressed in dB as follows:
\begin{equation}
PL_{UV,\text{A2G}}^{\text{$\mu$Wave}}\left(p_{UV}^{\text{LoS}}\right)=p_{UV}^{\text{LoS}}PL_{UV}^{\text{LoS}}+(1-p_{UV}^{\text{LoS}})PL_{UV}^{\text{NLoS}},
\end{equation}
where $p_{UV}^{\text{LoS}}$ is the probability of having a LoS link between the nodes $U$ and $V$ and is expressed as follows~\cite{PL3}:
\begin{equation}
p_{UV}^{\text{LoS}}=\frac{1}{1+\nu_1 \exp(-\nu_2[\theta(\Delta_{UV})-\nu_1])},
\end{equation}
where $\theta(\Delta_{UV})$ is the elevation angle between nodes $U$ and $V$ in degree. The constants $\nu_1$ and $\nu_2$ depend on the environment, e.g., urban/non urban and density of buildings and their altitudes. Notice that, when the drones are located at high altitude, the probability of having a LoS becomes higher.

\subsubsection{Path Loss Model for the mmWave Band}
The mmWave offers a large amount of free spectrum (i.e., around 100 GHz) with large bandwidths allowing extremely high throughput compared to $\mu$Wave bands but requires highly direction links to cope with increased path loss. We distinguish three bands in mmWave: i) the V-band (57 to 70 GHz) and above, ii) the E-band (70 GHz and less than 86 GHz) and finally, iii) the D-band (110-170 GHz). This enables the use of bandwidths higher than at least 500 MHz. However, mmWave links are suffering from several issues that can significantly attenuate the signals. Indeed, in addition to the free-space path loss effect, the signal is affected by atmospheric conditions such as oxygen, vapor, and rain. The International Telecommunication Union (ITU) has modeled the attenuation in dB at distance $\Delta_{UV}$ due atmospheric conditions as follows~\cite{mesodiakaki2016energy}:
\begin{equation}
PL_{UV}^\text{Atm}=\frac{\Delta_{UV}}{1000}\left(L_{vap}+L_{O_2}+L_{rain}\right),
\end{equation}
where $L_{vap}$, $L_{O_2}$, and $L_{rain}$ correspond to signal attenuation due to vapor water, oxygen, and rain and their expressions can be obtained from~\cite{mesodiakaki2016energy}. Furthermore, mmWave signals cannot penetrate solid materials very well therefore, LoS links are mandatory to enable mmWave communication. Hence, we consider that mmWave can be used for all A2A links and A2G links if $p_{UV}^{\text{LoS}}\approx 1$. Hence, the total average path loss of mmWave links in dB can be written as follows:
\begin{subequations}
\begin{align}
&PL_{UV,A2A}^\text{mmWave}=PL_{UV}^\text{Atm}+PL_{UV}^{\text{LoS}},\\
&PL_{UV,A2G}^\text{mmWave}=\left\{
   \begin{array}{ll}
	PL_{UV}^\text{Atm}+PL_{UV}^{\text{LoS}}, & \text{if } p_{UV}^{\text{LoS}}\geq 1-\epsilon,\\
	-\infty, & \text{otherwise}.
	   \end{array}
                 \right.
\end{align}   
\end{subequations}          
where $\epsilon$ is a non-negative parameter close to zero.
\subsection{Drone Power Model}
It is important to efficiently manage the energy consumption of the drones as they are battery-limited. The total energy consumption $E^{\text{tot}}_d$ of a communicating drone $d$ is composed of two components, namely the hover and transition energies needed for its movement and the communication energy needed to relay the transceivers' data. The models of the hover power and the transition power levels of a drone $d$, denoted by $P^\text{hov}_d$ and $P^{\text{tr}}_d$, are given as~\cite{Dpower_model}:
\begin{equation}\label{Ps}
P^\text{hov}_d= \sqrt{\frac{(m_d g)^3}{2 \pi (r^\text{prop}_d)^2 n^\text{prop}_d \rho}}, \text{ and } P^{\text{tr}}_d=\frac{P^\text{full}_d}{v^\text{max}_d}v_d,
\end{equation}
where $m_d$, $r^\text{prop}_d$, and  $n^\text{prop}_d$ denote the mass in kg, the radius, and the number of the propellers of drone $d$, respectively. The earth gravity is denoted by $g$ while $\rho$ denotes the air density. The average drone's speed is denoted by $v_d$ and is assumed to be constant during the trip. Finally, $v^\text{max}_d$ is the maximum speed of the drone while $P^\text{full}_d$ denotes the power required to move the drone at its full speed.

On the other hand, the power consumption of a drone $d$ serving the pair of transceivers $n$ due to data transfer can be approximated by a linear model as follows~\cite{EARTH}:
\begin{align}\label{BSpowermodel}
\small   P^{\text{com}}_{nd}=&P^{\text{static}}_d\hspace{-0.06cm}+\hspace{-0.06cm}\pi_{nd} \alpha_{d}^{\text{$\mu$Wave}}\hspace{-0.06cm}P_{d}^{\text{$\mu$Wave}}\hspace{-0.06cm}+\hspace{-0.06cm}(1-\pi_{nd} )\alpha_{d}^{\text{mmWave}}\hspace{-0.06cm}P_{d}^{\text{mmWave}},
\end{align}
where $P^{\text{static}}_d$ is the static power part of the communication component of drone $d$ while, $\alpha_{d}^{\text{$\mu$Wave}}$ and $\alpha_{d}^{\text{mmWave}}$ are parameters that scales with the radiated power for the $\mu$Wave and mmWave bands. The binary parameter $\pi_{nd}$ indicates whether pair $n$ and drone $d$ are using the $\mu$Wave or mmWave band ($\pi_{nd}=1$ if $\mu$Wave band used and $\pi_{nd}=0$ otherwise). It should be noted that using both technologies at the same time by a pair of transceivers is useless since if it is possible to use the mmWave band, the throughput due to $\mu$Wave band can be then neglected. In the opposite case, the use of $\mu$Wave band systematically indicates that the mmWave channel is not available. The number of drones using the $\mu$Wave can be then calculated as $\mathcal N^\text{$\mu$Wave}=\sum_{d=1}^{D}\sign\left(\sum_{n=1}^N\pi_{nd}\right)$ where $\sign(.)$ is the sign function. Finally, we denote by $\bar{E}_d$ and $\bar{B}_d$ the energy consumption and the battery capacity of drone $d$.

\section{Problem Formulation}
\label{Sec3}
The objective of this framework is to minimize the total service time of the network, in other words, the time needed to transmit the $N$ messages requiring relaying support. The data transmission can be performed in sequential and parallel ways. Sequentially means that two pairs will be served by the same drone. In this case, the drone serves one pair of transceivers then, moves to serve another one. Data transmission can be also performed in a parallel manner when multiple drones are used simultaneously to serve different pairs of transceivers. The service time depends on the types of the used band, i.e., $\mu$Wave or mmWave, the distance separating the nodes, and the positions of the drone stops. 

We denote by $S_{nd}$ the service time needed to serve a pair of transceivers $n$ by a drone $d$. It includes the time needed to arrive at location $\boldsymbol{X_{nd}}$ at which it will serve the pair $n$ plus the communication time needed to transfer the message of size $M_n$. The communication time depends on the used communication band. Its expression is given as follows:
\begin{align}
\label{comtime}
T^c_{nd}=\pi_{nd}  \frac{M_n}{R^{\text{$\mu$Wave}}_{nd}}+(1-\pi_{nd} )\frac{M_n}{R^{\text{mmWave}}_{nd}},
\end{align}
where i) the first term in~\eqref{comtime} indicates the communication time due to the support of drone $d$ using the $\mu$Wave band. The associated data rate using the DF strategy $R^{\text{$\mu$Wave}}_{nd}$ is then given as follows:
\begin{align}
R^{\text{$\mu$Wave}}_{nd}=\frac{B^{\text{$\mu$Wave}}}{2\mathcal N^{\text{$\mu$Wave}}}\min\big(&\log_2\left(1+\text{SINR}^{\text{$\mu$Wave}}_{T_nd}\right),\notag\\
&\log_2\left(1+\text{SINR}^{\text{$\mu$Wave}}_{dR_n}\right)\big),
\end{align}
where $\min(.,.)$ denotes the minimum function. Recall that, in DF relaying, the transmission is made over two time slots.
\\
ii) the second term in~\eqref{comtime} measures the communication time when the data is transmitted via the support of drone $d$ over the mmWave band. The data rate $R^{\text{mmWave}}_{nd}$ is then given as:
\begin{align}
R^{\text{mmWave}}_{nd}=\frac{B^{\text{mmWave}}}{2}\min\big(&\log_2\left(1+\text{SNR}^{\text{mmWave}}_{T_nd}\right),\notag\\
&\log_2\left(1+\text{SNR}^{\text{mmWave}}_{dR_n}\right)\big).
\end{align}

The service time $S_{nd}$, whose expression is given in~\eqref{ServiceTime_n}, starts when the drones leave the DS ($m=0$). Hence, we assume that $S_{0d}=0, \forall d=1,\dots,D$. 
\begin{equation}
\label{ServiceTime_n}
\small S_{nd}=\sum_{m=0\atop m\neq n}^{N} p_{mnd} S_{md}+\sum_{m=0\atop m\neq n}^{N}  p_{mnd}T^f_{mnd}+ T_{nd}^c\left(\sum_{m=0\atop m\neq n}^{N} p_{mnd}\right).
\end{equation}
In~\eqref{ServiceTime_n}, the first term indicates the service time of the previous pair $m$ served by drone $d$. The second term in~\eqref{ServiceTime_n} measures the time spent by drone $d$ to move from location $\boldsymbol{X}_m$ to location $\boldsymbol{X}_n$ where $m=0,\dots,N$ and $n=1,\dots,N$, and finally, the last term measures the communication time to complete the data transfer.
In~\eqref{ServiceTime_n}, $p_{mnd}$ is a binary variable that equals 1 if drone $d$ is directly serving pair $n$ after pair $m$. It is introduced to indicate the path of each drone starting from DS, passing by the stops till returning back to the DS. The parameter $T^f_{mnd}$ denotes the flying time needed by drone $d$ to move from the location at which it was serving the pair $m$ to the location at which it is serving pair $n$. It is calculated according to the speed of the drone as $T^f_{mnd}=\frac{\Delta_{mn}}{v_d}$. For simplification, we assume that the drones adapt a rectilinear motion in the 3D space.

The service time of a pair of transceivers $n$ is then denoted $S_n$ and is expressed as follows:
\begin{equation}
S_n=\sum_{d=1}^{D}\left(\sum_{m=0}^{N} p_{mnd}\right)S_{nd}.
\end{equation}
The total energy consumption accounting the flying, hovering, and communication energies is expressed as follows:
\begin{align}
E_d=&\left(P^{\text{hov}}_d+P^{\text{tr}}_d\right)\sum_{m=0}^{N}\sum_{n=0}^N p_{mnd} T^f_{mnd} \notag\\
&+\sum_{n=1}^N \left(P^{\text{hov}}_d+P_{nd}^{\text{com}}\right)T^c_{nd}\left(\sum_{m=0}^N p_{mnd}\right).
\end{align}
Note that the energy expression above takes into account the energy needed by a drone to fly back to the DS. The optimization problem minimizing a weighted sum of the service times of all pairs of transceivers requiring drones' support using the $\mu$Wave or mmWave band is given as follows\footnote{For readability, we use the following notations $\forall d$, $\forall m$, and $\forall n$ to denote $\forall d=1,\dots, D$, $\forall m=0,\dots, N$, and $\forall n=1,\dots, N$, respectively. Otherwise, the range of each index will be specified.}:
\begin{subequations}\label{ProblemP}
\begin{align}
\hspace{-0.3cm}\text{(P):}\;&\underset{\pi_{nd} \in\{0,1\},p_{mnd}\in\{0,1\},\atop \boldsymbol{X}_n\in \Omega,}{\text{minimize}} \quad\quad \sum_{n=1}^N\omega_nS_{n} \label{obj}\\   
&\text{subject to:}\notag\\
&\sum_{m=0\atop m\neq n}^N p_{mnd}\leq 1,\;\;\forall n,\forall d, \sum_{n=1\atop n\neq m}^N p_{mnd}\leq 1,\;\;\forall m,\forall d,\label{const1}\\
&p_{mnd}+p_{nmd}\leq 1,\;\;\forall m=1,\dots, N, \forall n,\forall d,\label{const2}\\
&\sum_{m=1\atop n\neq m}^N p_{nmd}\leq \sum_{m=0\atop m\neq n}^N p_{mnd},\;\;\forall n,\forall d,\label{const3}\\
&\sum_{m=1\atop n\neq m}^N p_{m0d}=1,\;\;\forall d,\label{const4}\\
&\sum_{m=1\atop n\neq m}^N \sum_{d=1\atop n\neq m}^D p_{mnd}=1,\;\;\forall d,\label{const5}\\
&E_d\leq \bar{B}_d,\;\;\forall d,\label{const6}
\end{align}
\end{subequations}
where i) $\pi_{nd}$: binary variable indicating which technology is used for data transmission, ii) $p_{mnd}$: binary variable identifying the path of drone $d$ (i.e., $p_{mnd}=1 $, if drone $d$ supports pair $n$ directly after pair $m$), and iii) $\boldsymbol{X}_n$: $3\times 1$ continuous variable determining where a drone will be located to serve the pair of transceivers $n$, are the decision variables of the problem (P). It is noteworthy that the values of $\boldsymbol{X}_n$ have a direct impact on the data rate $T^c_{nd}$ as well as the distances to be traveled by the drone and hence, on the service time $S_n$. 

In (P), we propose to minimize a weighted sum of the different transceivers' service times ($S_1,\dots,S_N$). The values of the parameters $\omega_n$ where $\omega_n\in [0,1]$ and $\sum_{n=1}^N\omega_n=1$, are chosen by the network operator. They can be used to promote some pairs of transceivers by giving them more priority to be served first or they can be used for fairness purpose, e.g., give priority to nodes that have a higher message size. Otherwise, their values can be set uniformly as $\omega_n=\frac{1}{N}$.
 
Constraints~\eqref{const1} force a drone to go to at maximum one destination when leaving its current location. Constraints~\eqref{const2} prohibit a drone to return to a node that it just left. Together, constraints~\eqref{const1} and~\eqref{const2} avoid having cyclic operation of the drones between multiple pairs of transceivers. Constraints~\eqref{const3} ensure that a drone can leave a location $\boldsymbol{X}_n$ at which it serves the pair of transceivers $n$ only if it already arrived there. Constraints~\eqref{const4} force the drone to return to DS after completing its tour, while constraints~\eqref{const5} indicate that each pair of transceivers must be served by at maximum one drone. Finally, constraints~\eqref{const6} ensure that the energy consumption due to the motion of drone $d$ and data transmission must not exceed its energy budget. This includes the energy needed to return to the DS. It is worth to note that if $\left(\sum_{n=1}^N\sum_{m=0\atop m\neq n}^N p_{mnd}=0\right)$ then, we can deduce that drone $d$ is not serving any pair $n$ and remains in DS.

The optimization problem (P) is classified as a mixed integer non linear programming problem where its optimal solution is difficult to obtain. Therefore, in the next section, we propose an iterative algorithm aiming at finding an efficient solution for drones' management. The problem can be infeasible if the energy budget of all drones is not sufficient to serve all the transceivers nodes. Therefore, it is recommended that the operator provides sufficient resources, e.g., energy and drones, to be able to serve all the nodes.

\section{Proposed Iterative Solution}
\label{Sec4}
The proposed solution is an iterative approach aiming to determine i) the band to be used by the pair of transceivers identified by $\pi_{nd}$, ii) the location $\boldsymbol{X}_n$ where a drone needs to statically hover to relay the data of the pairs of transceivers, and iii) the path of each drone $p_{mnd}$. We proceed by a four-step algorithm:

$\bullet$ \textbf{Step 1}: 3D potential relaying locations:
In this step, that we identify by the iteration $t=0$, the objective is to determine where a drone $d$ can be potentially located to serve a pair of transmitter $n$. This is determined by finding two locations, denoted by $\boldsymbol{X}_{nd}^\text{$\mu$Wave}$ and $\boldsymbol{X}_{nd}^\text{mmWave}$, maximizing the total throughput if the drone $d$ is using the $\mu$Wave and the mmWave bands, respectively. In other words, we aim to solve the following unconstrained non-convex problem (P0) for each band $\mathcal X \in \{\text{$\mu$Wave},\text{mmWave}\}$ and drone $d$:
\begin{align}
\text{(P0):}\;&\underset{\boldsymbol{X}_{nd}^{\mathcal X}\in \Omega}{\text{maximize}} \; R^{\mathcal X}_{nd}, \;\; \forall n, \forall d, \text{ with }\mathcal X \in \{\text{$\mu$Wave},\text{mmWave}\}.\label{obj0}
\end{align}
The above optimization problem can be solved using numerical or meta-heuristic algorithms such as the Newton method or particle swarm optimization. Its output will provide initial locations for the drones to be employed in the next steps of the proposed approaches. We denote them by $\boldsymbol{X}_{nd}^\text{$\mu$Wave}(0)$ and $\boldsymbol{X}_{nd}^\text{mmWave}(0)$. The problem is solved for each drone since the drone may have different characteristics such as the transmit power levels. So, different locations might be obtained. Note that, for the $\mu$Wave band and for fair comparison, the throughput in (P0) is determined assuming full use of the bandwidth for all drones and pairs of transceivers.

The solution of these problems also allow to determine whether a drone $d$ will utilize the mmWave band when relaying the data or not. Indeed, in some cases, e.g., the transceivers are located far from each other or when a LoS link cannot be established between the drone and one of the nodes, the numerical method will not converge and hence, the drone is able to use the $\mu$Wave band only, i.e., $\pi_{nd}=1$. Otherwise, the mmWave will be used and hence, the parameter $\pi_{nd}=0$.

$\bullet$ \textbf{Step 2}: Paths of the drones:
Once initial locations $\boldsymbol{X}_{nd}^{\mathcal X}(0)$ and the bands to be used $\pi_{nd}$ are determined in \textbf{Step 1}, we aim at \textbf{Step 2} to assign the drones to the pairs of transceivers and determine their paths. To do so, we convert the optimization (P) into a MILP that determines the values of the decision variables $p_{mnd}(0)$ and hence, the service time of each drone given the locations $\boldsymbol{X}_{nd}^{\mathcal X}(0)$. Indeed, the service time of each pair of transceiver $n$, $S_n$, depends on the chosen path of the drone $d$ as given in~\eqref{ServiceTime_n}. Therefore, we assume that $S_{nd}$ is a decision variable and we convert (P) to a MILP by linearizing the product of binary and continuous variables $p_{mnd}S_{md}$ in \eqref{ServiceTime_n} by introducing a new decision variable denoted by $\tilde{S}_{mnd}$ such that $\tilde{S}_{mnd}=p_{mnd}S_{md}$. The following linear constraints are then added to the problem:
\begin{subequations}
\label{LinearProd}
\begin{align}
&\tilde{S}_{mnd}\leq \bar{S}_{md}p_{mnd}, \forall m, \forall n, \forall d,\\
&\tilde{S}_{mnd}\leq S_{md}, \forall m, \forall n, \forall d,\\
&\tilde{S}_{mnd}\geq S_{md}-\left(1-p_{mnd}\right), \forall m, \forall n, \forall d,\\
&\tilde{S}_{mnd}\geq 0,\forall m, \forall n, \forall d,
\end{align}
\end{subequations}
where $\bar{S}_{md}$ is an upper bound of $S_{md}$ and can be a sufficiently large positve number. Hence, the MILP, denoted by (P1), can be written as follows:
\begin{align}
\text{(P1):}&\;\underset{p_{mnd},S_{nd},\tilde{S}_{mnd}}{\text{minimize}} \quad\quad  \sum_{n=1}^N\omega_n\sum_{d=1}^DS_{nd}(0) \label{obj1}\\   
&\text{subject to:}\notag\\
&\hspace{-1cm}S_{nd}\hspace{-0.06cm}=\hspace{-0.1cm}\sum_{m=0\atop m\neq n}^{N} \tilde{S}_{mnd}\hspace{-0.06cm}+\hspace{-0.06cm}\sum_{m=0\atop m\neq n}^{N}  p_{mnd}T^f_{mnd}\hspace{-0.06cm}+\hspace{-0.06cm} T_{nd}^c\left(\sum_{m=0\atop m\neq n}^{N} p_{mnd}\right),\\
&\eqref{const1}, \eqref{const2}, \eqref{const3}, \eqref{const4}, \eqref{const5},  \eqref{const6}, \text{and } \eqref{LinearProd}. \notag
\end{align}

The optimization problem (P1) is a MILP and can be solved optimally using off-the-shelf software. The outputs of (P1) identify the paths of each drone $d$ given the locations $\boldsymbol{X}_{nd}^{\mathcal X}(0)$ determined in \textbf{Step 1}. Consequently, the drone serving each pair of transceivers is known and the service time is computed. 

$\bullet$ \textbf{Step 3}: Adjustment of the drone stops: A path for each drone is determined in \textbf{Step 2}. Nevertheless, this path can still be improved such that the total service time is reduced. Indeed, in \textbf{Step 1}, the location of a drone $d$ is selected such that the throughput of pair $n$, $R_{nd}^\mathcal X$, is maximized and hence, the time needed to transfer the message, $T^c_{nd}$, is minimized. However, in some cases, slightly modifying the location of the drone may increase the communication time but, at the same time, helps in reducing the flying time of the drone and hence, the service time of the pair $n$ itself and the rest of the pairs of transceivers served by the drone $d$. Therefore, we aim in \textbf{Step 3} to adjust the stops of each drone such that the total service time is improved. Since the drone stops are inter-dependent, we propose to deal with each path, i.e., drone, separately and proceed with a 3D  hierarchical search centered around $\boldsymbol{X}_{nd}(0)$. The objective of the algorithm is to solve the following optimization problems (P2) for each drone:
\begin{align}
\text{(P2):}&\;\underset{\boldsymbol{X}_{nd}}{\text{minimize}} \quad\quad  \sum_{n=1}^N\omega_nS_{nd}(0), \quad\quad\quad\forall d, \label{obj2}\\   
&\text{subject to: }\quad\;\eqref{const6}.\notag
\end{align}

\begin{figure}[t!]
  \centerline{\includegraphics[width=7.8cm]{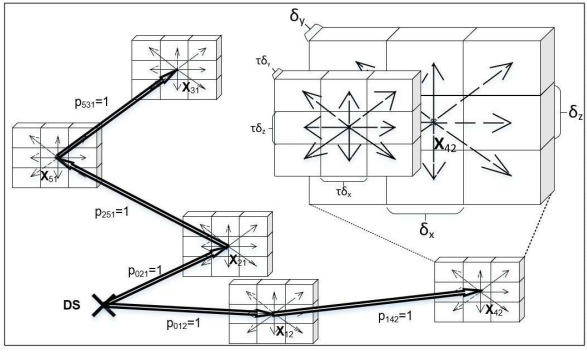}}\vspace{-0.4cm}
   \caption{\, Examples of 3D hierarchical cubical search for two drones given their paths. In this example, the search is made over the x and z axes.    \normalsize}\label{Ritning1}\vspace{-0.4cm}
\end{figure}

To this end, the 3D hierarchical search assumes that the locations of the drone $\boldsymbol{X}_{nd}$ can be shifted locally in a certain number of directions according to the accuracy of the search. For instance, a cubical search can be employed. Hence, $26$ directions can be checked to form a rectangular Cuboid of size $3\delta_x \times 3\delta_y \times 3\delta_z$ where $\delta_x$, $\delta_y$, and $\delta_z$ are the shifting distances from $\boldsymbol{X}_{nd}$ over the x-axis, y-axis, and z-axis, respectively. An illustration of the 3D cubical search over the x and z axes is given in Fig.~\ref{Ritning1}. The shifted position of a drone cannot exceed the boundaries of $\Omega$. Hence, the minimum/maximum possible location is selected. It may also be possible that, when shifting a mmWave drone, the LoS link is lost. In this case, this location is eliminated from the cubical search.

Once the possible shifting locations are determined for each drone stop, a computation of the total service time is made for all the possible combinations. The locations offering the lowest service time and that do not violate the energy budget constraint are then selected. Hence, for each drone stop, one of the $27$ possible locations (26 potential new locations + the current one) is selected. Then, the dimension of the Cuboid is decreased by a coefficient $\tau$ so the search is made over a small cuboid centered around the new optimized locations. This process is repeated until no enhancement is reached. The final locations represent the updated stops of drone $d$.

The 3D cubical search aims to find a better path minimizing the total service by essentially reducing the flying time of the drone while maintaining connectivity. It looks for all possibilities defined by the hierarchical search. It requires a high computational cost but remains a simple approach that guarantees a high accuracy. Nevertheless, it is possible to reduce the number of search locations for reduced accuracy and faster convergence. The complexity does not impose a significant concern as we are dealing with a proactive approach optimizing the drones' trajectories once.

$\bullet$ \textbf{Step 4}: Iterate until convergence: In this step, we proceed with a series of iterations between \textbf{Step 2} and \textbf{Step 3} until convergence is reached. The convergence is reached when the objective function given in~\eqref{obj} is not improved by more than $\upsilon$ where $\upsilon\geq 0$. In Step 4, the paths of certain drones may change. This case usually happens when some drones are serving close pairs of transceivers. A flowchart of the proposed iterative approach for flying cooperative $\mu$Wave/mmWave drone is given in Fig.~\ref{Flowchart}.

\begin{figure}[t!]
\vspace{0.25cm}
  \centerline{\includegraphics[width=5.75cm]{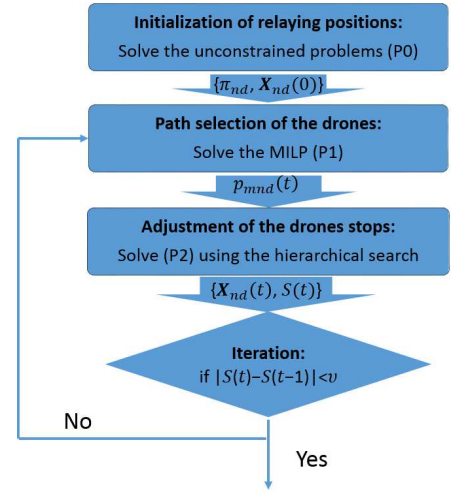}}\vspace{-0.4cm}
   \caption{\, Flowchart summarizing the proposed algorithm.    \normalsize}\label{Flowchart}\vspace{-0.25cm}
\end{figure}

\section{Results and Discussion}
\label{Sec5}
In this section, we investigate the impact of some parameters on the system performance. We consider a bounded area of size $5\times 5 \times 0.2$ km$^3$ where the DS is located at the center of the area $\boldsymbol{X}_0=(2500, 2500, 30)$. Without loss of generality, we set the simulation parameters as given in Table~\ref{tab2}~\cite{EARTH,Dpower_model,mesodiakaki2016energy}. The transmitters are randomly placed in the region of interest following a uniform distribution. Their corresponding receivers are located at a random distance between $0.3$ and $2$ km from them. In the simulations, we consider that all the transceivers are ground nodes. The hierarchical cuboid search is done with the following parameters: $\delta_x=\delta_y=300$ m, $\delta_z=50$ m, and $\tau=0.6$.

{\small
\begin{table}[t]
\centering
\caption{\label{tab2} System parameters}
\vspace{-.15cm}
\addtolength{\tabcolsep}{-4pt}\begin{tabular}{|l|c||l|c|}
\hline
\textbf{Parameter} & \textbf{Value} & \textbf{Parameter} & \textbf{Value}\\ \hline \hline
$B^{\text{$\mu$Wave}}$ (MHz) & 1 & $P_{T_n}^{\text{mmWave}}=P_{d}^{\text{mmWave}}$ (dBm) & 24 \\ \hline
$B^{\text{mmWave}}$ (GHz) & 3.5 & $P_{T_n}^{\text{$\mu$Wave}}$, $P_{d}^{\text{$\mu$Wave}}$  (dBm) & $[-10,36]$ \\ \hline
$m_{\text{tot}}$ (kg) & 2 & $G_{T_n}^{\text{$\mu$Wave}}=G_{R_n}^{\text{$\mu$Wave}}=G_{d}^{\text{$\mu$Wave}}$ (dBi) & 0 \\ \hline
$r_d^\text{prop}$ (cm) & $20$ & $G_{T_n}^{\text{mmWave}}=G_{R_n}^{\text{mmWave}}=G_{d}^{\text{mmWave}}$ (dBi) & 37  \\ \hline  
$n_d^\text{prop}$ & 4 & $M_n$ (MB) & $[250,625]$\\ \hline 
  $P_{\text{full}}$ (W) & 10 & $v_d=v_d^{\text{max}}$ (m/s) & $[10,20]$\\ \hline
\end{tabular}\vspace{-0.4cm}
\end{table}
}

\begin{figure}[t!]
\vspace{0.2cm}
\centerline{\includegraphics[width=8.5cm]{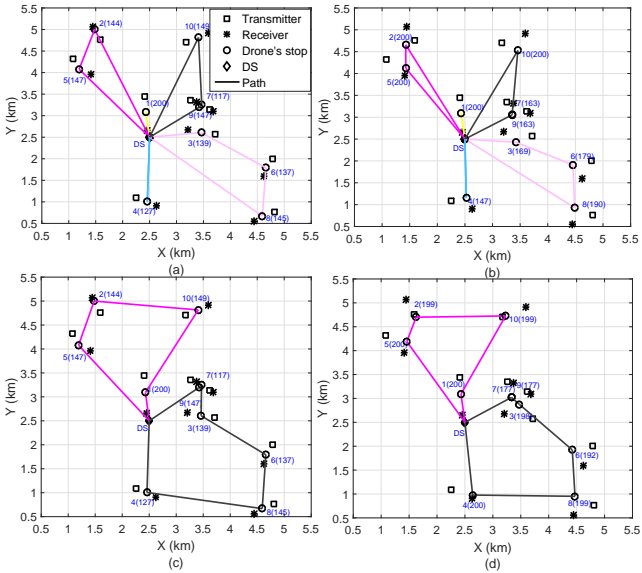}}\vspace{-0.3cm}
\caption{\, Optimized trajectories (a) $D=5$ using MILP only, (b) $D=5$ using MILP and cubical search, (c) $D=2$ using MILP only, and (d) $D=2$ using MILP and cubical search (Notation x(y): pair's index (drone's altitude)).   \normalsize}\label{Fig3}\vspace{-0.0cm}
\end{figure}

\begin{figure}[t!]
\centerline{\includegraphics[width=7.5cm]{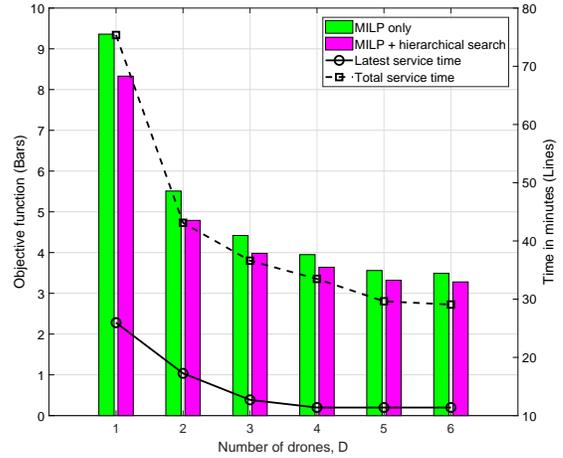}}\vspace{-0.4cm}
\caption{\, Performance of the proposed scheme for a scenario with $N=10$ pairs of transceivers with $\mathcal N^\text{$\mu$Wave}=1$ and $w_{n}=\frac{M_n}{\sum_{n=1}^NM_n}$.   \normalsize}\label{Fig4}\vspace{-0.45cm}
\end{figure}

In Figs.~\ref{Fig3} and~\ref{Fig4}, we consider identical drones with fixed transmit power levels $23$ dBm and fixed speed $15$ ms$^{-1}$ and sufficiently large charged battery. We then vary the number of used drones from 1 to 6 for a fixed scenario where $10$ pairs of transceivers are requiring relaying support. Pair $1$ is using the $\mu$Wave band while the rest of the pairs are using the mmWave band. In Fig.~\ref{Fig3}, we select two cases, i.e., $D=5$ and $D=2$ and plot the optimized trajectories using the MILP solution (Figs.~\ref{Fig3}(a) and (c)) and the MILP plus hierarchical search solution (Figs.~\ref{Fig3}(b) and (d)). When the MILP is used, the drones' stops correspond to the best locations at which the communication time is minimized. Hence, the drones' stops are located between the transmitter and the receiver of each pair. The drones are then assigned to the pairs of transceivers optimally using the MILP. With $D=5$, one drone is assigned to the pair using the $\mu$Wave as it requires a high communication time, and one drone is assigned to pair $4$ as it is located quite isolated from the other nodes. When the number of drones is reduced to $D=2$, each of these nodes are assigned to different drones to reduce the total service time. Hence, the $\mu$Wave band is assigned to drone $1$ while pair $4$ is assigned to the other drone. When the hierarchical search is employed, the locations of the drones' stops are modified such that the flying time is reduced and the areas of the polygons representing the paths of the drones are decreased. A tradeoff between the flying time and the communication time is then obtained. The altitudes of the drones increase to ensure the establishment of LoS links while shifting their locations.

In Fig.~\ref{Fig4}, we plot the achieved weighted objective function using both solutions. The gain due to the hierarchical search is significant especially when using a small number of drones. In the right axis, we plot the total of service time corresponding to the sum of all the service times achieved by the drones. Increasing the number of drones significantly reduce the total service time. However, it is noticed that with 5 drones we are able to achieve close results to the case with 6 drones. Indeed, with $D=6$, one drone is exclusively assigned to pair $8$ which requires a significant flying time. Associating this pair to the drone serving pairs $3$ and $6$ as with the case of $D=5$ will almost achieve the same result. Hence, it is important to determine the necessary number of drones to serve the dispersed nodes. The latest service time representing the latest time needed to serve a pair of transceiver decreases with the increase of the number of drones. However, it stagnates starting from $D=3$ since starting from this value, only one drone is assigned to the $\mu$Wave pair and hence, its service time is dominated by its communication time. With $D=1$, this pair requires around 25 minutes to complete its transmission as the drone decides to serve the mmWave pairs first.

\section{Conclusion}
\label{Sec6}
This paper provides a generic framework to employ a UAV swarm as flying relays for multiple pairs of transceivers. Designed for bandwidth hungry and delay-tolerant applications, two different bands the $\mu$wave and mmWave bands are exploited to transmit the data. Accordingly, the drones' trajectories are determined and optimized such that a weighted function of the total service time is minimized. In our ongoing work, we will focus on considering non-orthogonal transmission when using $\mu$wave and hence, extra coordination among UAVs is necessary to limit the interference effect.

\bibliographystyle{ieeetr}

\begin{thebibliography}{10}

\bibitem{Du2017}
J.~D. et~al., ``Gbps user rates using mmwave relayed backhaul with high-gain
  antennas,'' {\em IEEE J. Sel. Areas Commun.}, vol.~35, pp.~1363--1372, June
  2017.

\bibitem{Rappaport2013}
T.~S. Rappaport, S.~Sun, R.~Mayzus, H.~Zhao, Y.~Azar, K.~Wang, G.~N. Wong,
  J.~K. Schulz, M.~Samimi, and F.~Gutierrez, ``Millimeter wave mobile
  communications for {5G} cellular: {I}t will work!,'' {\em IEEE Access},
  vol.~1, pp.~335--349, May 2013.

\bibitem{7463007}
S.~Hayat, E.~Yanmaz, and R.~Muzaffar, ``Survey on unmanned aerial vehicle
  networks for civil applications: A communications viewpoint,'' {\em IEEE
  Commun. Surveys Tuts.}, vol.~18, no. 5, Fourth Quarter 2016.

\bibitem{PL3}
A.~Al-Hourani, S.~Kandeepan, and S.~Lardner, ``Optimal {LAP} altitude for
  maximum coverage,'' {\em IEEE Wireless Commun. Lett.}, vol.~3, pp.~569--572,
  Dec. 2014.

\bibitem{Rangan2014}
S.~Rangan, T.~S. Rappaport, and E.~Erkip, ``Millimeter-wave cellular wireless
  networks: {P}otentials and challenges,'' {\em Proc. IEEE}, vol.~102, no.~3,
  pp.~366--385, Mar. 2014.

\bibitem{Kong2017}
L.~Kong, L.~Ye, F.~Wu, M.~Tao, G.~Chen, and A.~V. Vasilakos, ``Autonomous relay
  for millimeter-wave wireless communications,'' {\em IEEE J. Sel. Areas
  Commun.}, vol.~35, pp.~2127--2136, Sept. 2017.

\bibitem{Li2016}
K.~Li, W.~Ni, X.~Wang, R.~P. Liu, S.~S. Kanhere, and S.~Jha, ``Energy-efficient
  cooperative relaying for unmanned aerial vehicles,'' {\em IEEE Trans. Mobile
  Comput.}, vol.~15, pp.~1377--1386, June 2016.

\bibitem{Fawaz2018}
W.~Fawaz, C.~Abou-Rjeily, and C.~Assi, ``{UAV}-aided cooperation for {FSO}
  communication systems,'' {\em IEEE Commun. Mag.}, vol.~56, pp.~70--75, Jan.
  2018.

\bibitem{Jeong2018}
S.~Jeong, O.~Simeone, and J.~Kang, ``Mobile edge computing via a {UAV}-mounted
  cloudlet: {O}ptimization of bit allocation and path planning,'' {\em IEEE
  Trans.Veh. Technol.}, vol.~67, pp.~2049--2063, Mar. 2018.

\bibitem{Zhang2018}
S.~Zhang, H.~Zhang, Q.~He, K.~Bian, and L.~Song, ``Joint trajectory and power
  optimization for {UAV} relay networks,'' {\em IEEE Commun. Lett.}, vol.~22,
  pp.~161--164, Jan. 2018.

\bibitem{Fan2018}
R.~Fan, J.~Cui, S.~Jin, K.~Yang, and J.~An, ``Optimal node placement and
  resource allocation for {UAV} relaying network,'' {\em IEEE Commun. Lett.},
  vol.~22, pp.~808--811, Apr. 2018.

\bibitem{Wu2018}
S.~Wu, R.~Atat, N.~Mastronarde, and L.~Liu, ``Improving the coverage and
  spectral efficiency of millimeter-wave cellular networks using
  device-to-device relays,'' {\em to appear in IEEE Trans. Commun.}, 2018.

\bibitem{Deng2017}
J.~Deng, O.~Tirkkonen, R.~Freij-Hollanti, T.~Chen, and N.~Nikaein, ``Resource
  allocation and interference management for opportunistic relaying in
  integrated mmwave/sub-6 {GH}z {5G} networks,'' {\em IEEE Commun. Mag.},
  vol.~55, pp.~94--101, June 2017.

\bibitem{He2017}
Z.~He, S.~Mao, S.~Kompella, and A.~Swami, ``On link scheduling in dual-hop
  60-{GH}z mmwave networks,'' {\em IEEE Trans. Veh. Technol.}, vol.~66,
  pp.~11180--11192, Dec. 2017.

\bibitem{mesodiakaki2016energy}
A.~Mesodiakaki, A.~Kassler, E.~Zola, M.~Ferndahl, and T.~Cai, ``{Energy
  efficient line-of-sight millimeter wave small cell backhaul: 60, 70, 80 or
  140 GHz?},'' in {\em IEEE Intl. Sympos. A World of Wireless, Mobile and
  Multimedia Netw. (WoWMoM 2016)}, Coimbra, Portugal, June 2016.

\bibitem{Dpower_model}
J.~V. Dries~Hulens and T.~Goedeme, ``How to choose the best embedded processing
  platform for onboard {UAV} image processing,'' in {\em Intl. Joint Conf.
  Comput. Vision, Imag. and Comput. Graphics Theory and Appl. (VISIGRAPP 2015),
  Berlin, Germany}, Mar. 2015.

\bibitem{EARTH}
{G. Auer et al.}, ``{D2.3 v2: {E}nergy efficiency analysis of the reference
  systems, areas of improvements and target breakdown}.'' {EARTH} ({E}nergy
  {A}ware {R}adio Ne{T}work Tec{H}nologies), {Tech. Rep.}, Jan. 2011.

\end{thebibliography}

\end{document}